# The Lightest 2D Nanomaterial: Freestanding Ultrathin Li Nanosheets by in-situ Electron Microscopy

Muhua Sun[1,6], Nore Stolte[2,6], Jianlin Wang[1], Jiake Wei[1], Pan Chen[1], Yu Zhao[1], Zhi Xu[1,4], Wenlong Wang[1,3,4*], Ding Pan[2,5*], Xuedong Bai[1,3,4*]

[1]Beijing National Laboratory for Condensed Matter Physics and Institute of Physics, Chinese Academy of Sciences, Beijing, China

[2]Department of Physics, Hong Kong University of Science and Technology, Hong Kong, China

[3]School of Physical Sciences, University of Chinese Academy of Sciences, Beijing, China

[4]Songshan Lake Materials Laboratory, Dongguan, Guangdong, China

[5]Department of Chemistry, Hong Kong University of Science and Technology, Hong Kong, China

[6]These authors contributed equally: Muhua Sun, Nore Stolte.

Corresponding Authors:

*Wenlong Wang,* wwl@iphy.ac.cn;

*Ding Pan,* dingpan@ust.hk;

*Xuedong Bai,* xdbai@iphy.ac.cn

**Abstract**

Lithium (Li) is the simplest metal and the lightest solid element. Here we report the first demonstration of controlled growth of two-dimensional (2D) ultrathin Li nanosheets with large lateral dimensions up to several hundreds of nanometres and thickness limited to just a few nanometres by in-situ transmission electron microscopy (TEM). The nanoscale dynamics of nanosheets growth were unravelled by real-time TEM imaging, which, in combination with density function theory (DFT) calculations indicates that the growth of *bcc* structured Li into 2D nanosheets is a consequence of kinetic control as mediated by preferential oxidization of the (111) surfaces due to the trace amount of $O_2$ (~$10^{-6}$ Pa) within TEM chamber. The plasmonic optical properties of the as-grown Li nanosheets were probed by cathodoluminescence (CL) spectroscopy equipped within TEM, and a broadband visible emission was observed that contains contributions of both in-plane and out-of-plane plasmon resonance modes.

**Introduction**

With the experimental discovery of graphene in 2004,[1] the flourish of two dimensional (2D) nanomaterial research has evoked ever-increasing interest also in 2D metals.[2-3] In principle, only those materials that possess inherent layered structures have the propensity to form 2D nanostructures because of the strong lateral chemical bonding in planes but weak van der Waals interaction between layers.[4] Metal solids, however, are inherently non-layered materials with metallic bonding in three dimensions, which makes the formation of free-standing 2D metals very difficult to achieve. Thus far, the wet chemistry methods have witnessed major advances in achieving the anisotropic 2D growth of ultrathin nanoplates and nanosheets of some noble metals (gold, palladium, rhodium, etc.) that exhibit pronounced plasmonic or catalytic properties.[2, 5-8]

As the lightest and simplest metal, lithium (Li) is one of the best representations of quasi-free electron models, such that its nanostructures may offer unique opportunities for testing and understanding of the dimensionality and size-dependence of physical properties variations in a fundamentally more simple system.[9-11] However, in reality, as a consequence of the extreme chemical reactivity of alkali metals, the well-established wet chemical methods that work well for noble metals are normally inapplicable to alkali metals. Here we report the first demonstration of controlled growth of free-standing 2D ultrathin Li nanosheets by utilizing an in-situ electrochemical platform inside a transmission electron microscopy (TEM). The high vacuum condition of TEM chamber provides a relatively ideal environment for the stable growth of Li nanosheets and real-time monitoring of the nanoscale dynamics of their growth process by in-situ TEM imaging.

The in-situ TEM electrochemical growth of Li nanosheets is accomplished by adopting a solid-state open-cell configuration with a piece of Li metal as the anode electrode, an individual carbon nanotube (CNT) as the cathode electrode, and the naturally grown thin lithium oxide ($Li_2O$) layer on Li metal as the solid-state electrolyte that electrically isolate the Li anode and CNT cathode while allowing $Li^+$ ions to pass through,[12-13] as shown in Fig. 1a and Fig. 1b. Upon applying a positive voltage (typically 3.5 V unless otherwise noted) on Li electrode with respect to CNT electrode, $Li^+$ ions are driven to transport through the $Li_2O$ layer and react with

electrons passed through the CNT counter electrode, according to the following half-cell reaction:

$$Li^+ + e^- \rightarrow Li^0 \qquad (1)$$

In supplementary Movie S1 and Figure 1c, we show the real-time morphological evolution of a representative Li nanosheet during growth. At the early growth period, a small-sized nanosheet is observed to form with the constituted apex forming angles of nearly 120°. Further anisotropic 2D growth proceeds through lateral extension, with the newly emerged edge constituting angles of 120° with their adjacent edges as well. Although the evolution of each facet seems quite uneven, a highly symmetrical hexagonal nanosheet finally formed anyway. By comparing the sequential TEM images frame by frame, it can be clearly seen that the contrast of the nanosheet remained invariant, indicating that the thickness of Li nanocrystals did not change during the whole growth process. Such thickness confined growth model, i.e., remaining thickness unchanged during growth, has been reported in the synthesis of ultrathin noble metal nanosheets and semiconductor CdSe nanoplates.[5, 14] Selected area electron diffraction (SAED) and electron energy loss spectroscopy (EELS) were conducted on the obtained nanosheet shown in the frame of 87s in Fig. 1c, confirming without ambiguity that the as-formed nanosheet is metallic Li (Fig. 1d and Fig. 1f). The diffraction pattern exhibits 6 bright and sharp spots in a 6-fold rotational symmetry corresponding to the $\{1\bar{1}0\}$ reflections of the body-cantered cubic (bcc) Li single-crystal orientated in the [111] direction, indicating that the top and bottom basal planes of the Li nanosheet are (111) planes; the crystallographic assignment for the well-defined side facets are $\{1\bar{1}0\}$ planes, as the structural model shown in Fig. 1e. The EELS spectrum in Fig. 1f shows several distinguishable peaks at 56.4, 65.2 and 72.6 eV, which are in good agreement with the Li-K edge (electron transition from 1s → 2p) of metallic Li reported previously.[15-16]

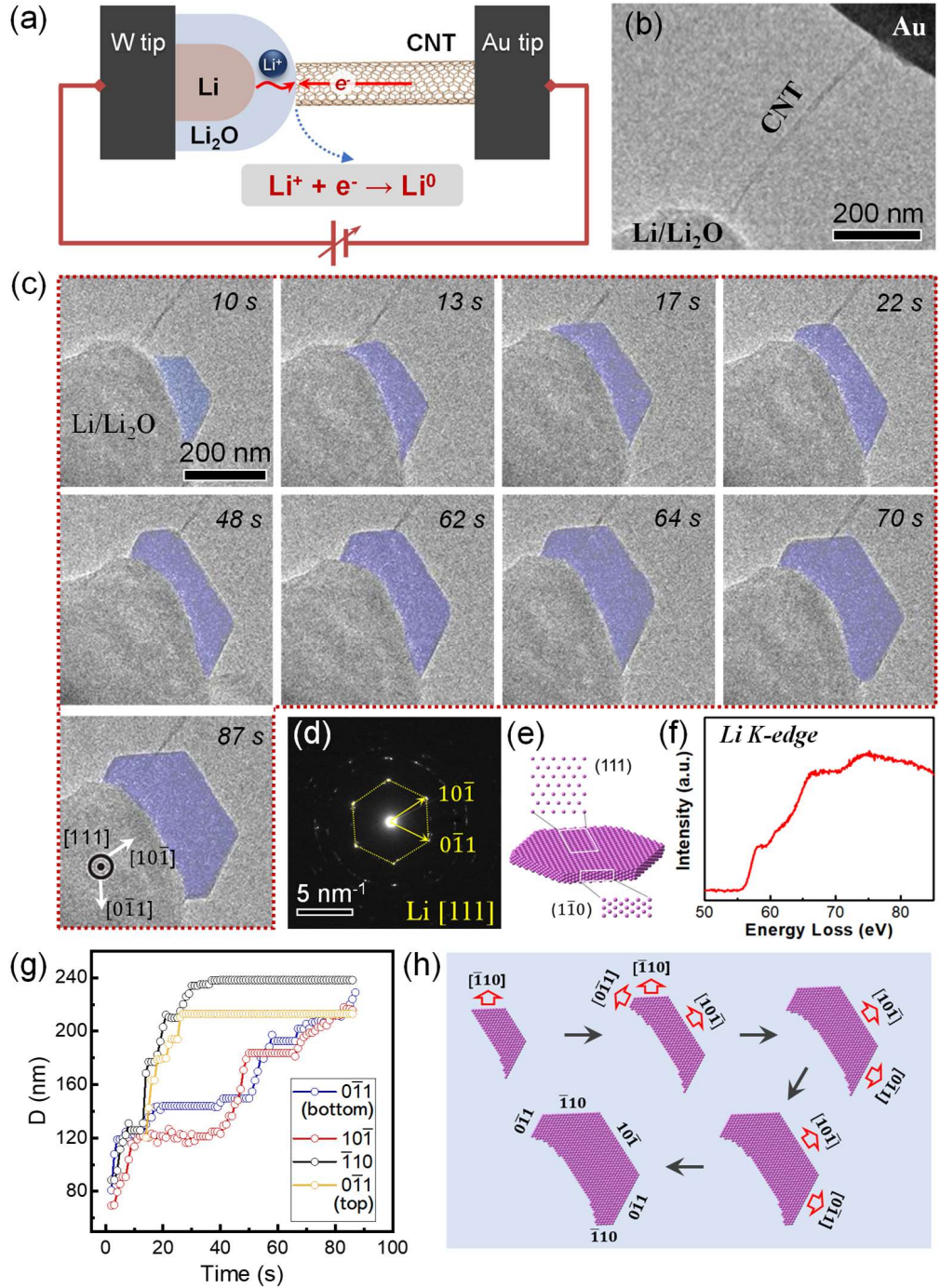

**Figure 1. Dynamic growth and detailed characterizations of 2D Li nanosheets by in-situ TEM.** Schematic (a) and the corresponding TEM image (b) of the experimental configurations for the electrochemical growth of metallic Li. (c) Time-lapsed TEM images showing the dynamic growth of Li nanosheet. The corresponding I-t curve recorded during the growth process is presented in Fig. S1. (d) SAED of the as-formed nanosheet shown in the frame of 87 s in (c). (e) Structural model of the well-faceted hexagonal Li nanosheet. (f) EELS spectrum collected from the newly formed nanosheet in (c). (g) The measured distances from the growing nanocrystal center to each of the facets as a function of time. (h) Schematic showing how the predominant growth of specific facets leads to the formation of a hexagonal nanosheet. The red arrows indicate the predominant growth directions.

To better understand the nanoscale kinetics of Li nanosheet growth, we quantified the facet development of the same Li nanosheet by tracking the evolution of different facets. The growing nanosheet sits along [111] direction during growth, which allows measurements of the distance from the centre of crystal to all the observable {1$\bar{1}$0} side faces. Changes in the distance from the center of crystal to the edges of each facet—(10$\bar{1}$), (0$\bar{1}$1) (bottom) , ($\bar{1}$10) and (0$\bar{1}$1) (top)—as a function of time are plotted in Fig. 1g. There are similar growth rates among these equivalent facets at first, while the subsequent growth of the facets fluctuates by presenting predominant growth at ($\bar{1}$10) and (0$\bar{1}$1) (top) facets then (10$\bar{1}$) and (0$\bar{1}$1) (bottom) facets. Although the growth fluctuation of the facets led to the formation of asymmetric nanosheet, the slow-growing facets can catch up, eventually forming a highly symmetrical hexagonal nanosheet. The schematics in Fig. 1h show how the predominant growth of specific facets leads to the formation of the subsequent nanostructures and finally to a hexagonal nanosheet, based on analysis of the in-situ TEM images in Fig. 1c. Growth fluctuations of equivalent facets may sound a bit counter-intuitive but are actually quite common in crystal growth when considered at the nanoscale. Similar growth fluctuations were also observed in the facet growth of platinum nanoparticles, where they found that the growth rates of four different {110} facets fluctuated due to the local environmental variations.[17]

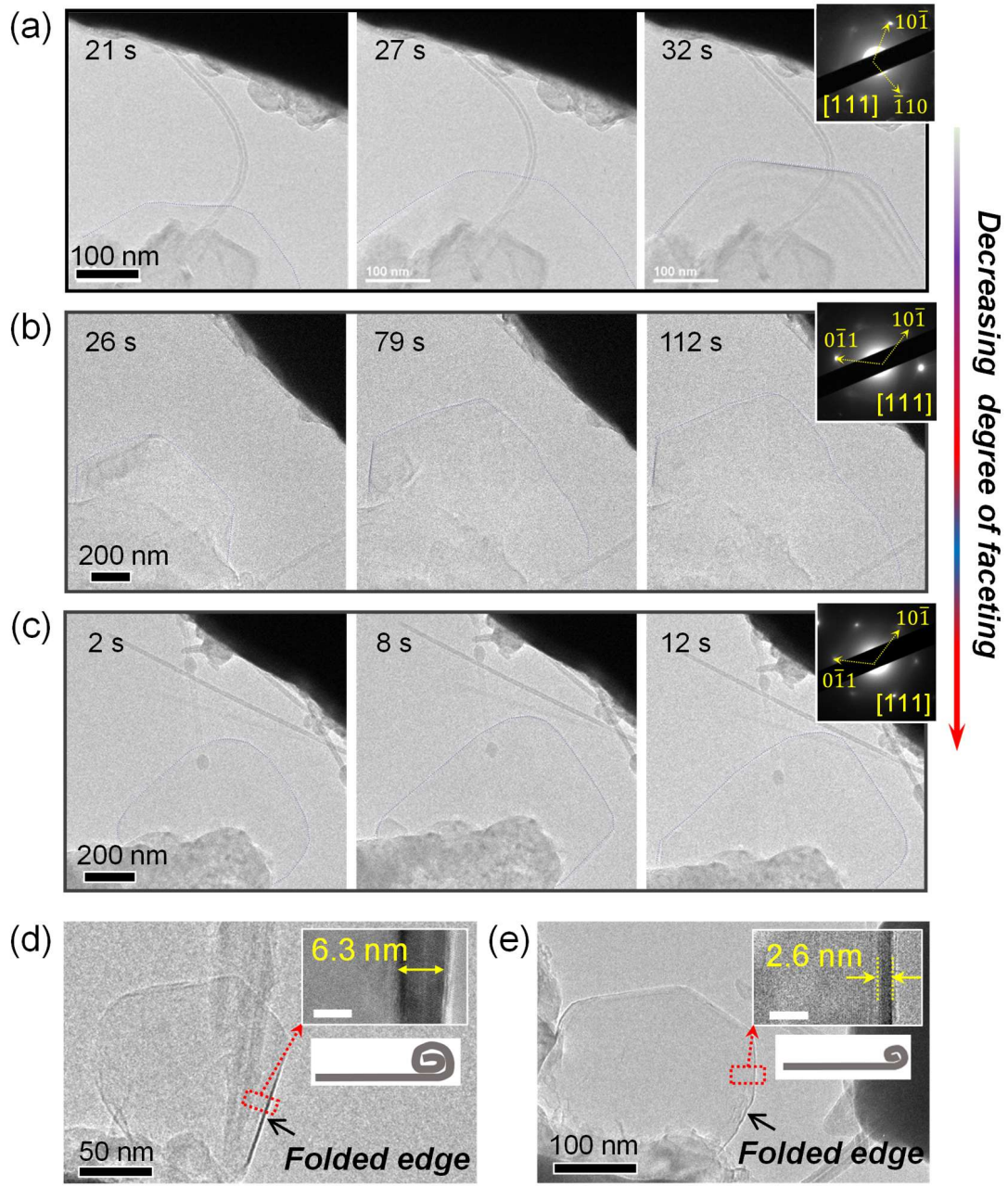

**Figure 2. 2D Li nanosheets with different degrees of faceting and the thickness measurement of the nanosheets.** (a)-(c) TEM images of the as-formed Li nanosheets that present decreasing degrees of faceting from the left to the right and the corresponding diffraction patterns. The models at the bottom depict the corresponding morphology of the as-formed nanosheets. (d)-(e) TEM image of the Li nanosheets that show folded edges. The insets show enlarged images of the red rectangle framed areas. The schematic diagrams illustrate how the edges are scrolled.

In our experiments, the growth rate of different nanosheets normally varies from sample to sample, even under the same applied voltage. It is only with relatively slow growth rate that a well-faceted nanosheet tend to be formed, whereas faster growth kinetics will constantly drive the system out of equilibrium and lead to the formation of nanosheets with less-faceted features. In Fig. 2a-2c (extracted from Movie S2-S4 respectively), we display the sequential TEM images of three Li nanosheets with decreased degree of faceting as compared to the one shown in Fig. 1c and movie S1. This observed correlation between growth rate and the morphology variations reflects the competition between faceting and coarsening during nanocrystal growth, and it is essentially a consequence of the equilibrium between the rates for atom deposition and surface diffusion.[18] In general, with relatively slower growth rate where the deposition of atoms onto the facets of the growing nanocrystals is slow, the adatoms are able to diffuse and migrate to the equilibrium position. In contrast, if the growth rate is fast, the deposition rate is much higher than the surface diffusion rate, the adatoms would mostly reside where they are deposited, leading to the coarsening of the growing surfaces.[19]

Interestingly, with a closer examination of the TEM images of the as-grown Li nanosheets, one can frequently see the occurrence of spontaneous folding of the nanosheet edges. Previous studies showed that the folding edge can be a measure to estimate the thickness of 2D nanostructures.[20-21] We show two nanosheets with the edges being folded differently (Fig. 2d-2e). The marked contrast difference between the folded edge and the unfolded area of the nanosheet in Fig. 2d indicates that the edge was multiply scrolled, as schematically illustrated by the inset image. In this circumstance, the real thickness of the nanosheet in Fig. 2d is certainly smaller than the width of the multiply scrolled edge which was measured to be 6.3 nm by higher resolution TEM image. For the nanosheet in Fig. 2e where the contrast difference between the folded edge and the flat area of the nanosheet is less distinguished, the higher resolution TEM image of the folded edge gives a thickness of ~ 2.3 nm. Furthermore, the thickness can also be estimated by comparing the integrated intensity of the zero-loss peak and that of the entire EELS spectrum (Fig. S3).[22-23] The above complementary thickness-measurement methods are able to draw a consistent and consolidated conclusion that the electrochemically grown Li nanosheets are quite thin in thickness that are less than 10 nm. Such thickness of the electrochemical grown Li nanosheets fall into quantum confined length scale, which rationalizes that the as-formed Li nanosheets are truly 2D metals.

As we have emphasized above, the electrochemical growth of 2D Li nanosheets takes advantage of the high vacuum condition in TEM chamber, so that the as-formed ultrathin Li nanosheets could escape from severe oxidation within a short time. However, the vacuum inside TEM chamber is not high enough for long-time maintenance of alkali metals in terms of their extremely high chemical reactivity. Based on the above considerations, the oxidation process of the as-formed Li nanosheet was investigated in details. The diffraction patterns of the as-formed Li nanosheet as a function of time are presented in Fig. S4, which exhibited the transformation from Li metal to mixed phases of $Li_2O$ and Li and finally into the pure $Li_2O$. The oxidation process of freshly formed metallic Li to $Li_2O$ is also confirmed by EELS comparisons of Li-K edge and O-K edge acquired from the newly formed nanosheet and the nanosheet kept in TEM chamber for 3h (see details in Fig. S4 and Fig. S5 in supporting information).

**Mechanistic understanding of 2D Li nanosheets growth**

Since Li is a bcc structured metal, the general ordering of surface energies of its three low index surfaces is: $\gamma_{110} < \gamma_{100} < \gamma_{111}$.[24] In our present work, the as-grown Li nanosheets expose dominant surfaces of (111) basal planes that possess higher surface energies, representing a thermodynamically unfavorable nanostructure. This suggests that the formation of 2D Li nanosheets is a consequence of kinetic-control growth pathways where the selective surface passivation of different facets may play a crucial role.[25] This conceptual framework is on the basis of previous experimental and theoretical studies of the shape-controlled synthesis of colloidal nanocrystals via wet-chemistry methodologies, where the selective passivation by structure-directing agents is a well-established mechanism that accounts for anisotropic growth of various metals and semiconductor nanostructures, including 2D ultrathin nanoplates and nanosheets.[5, 26-27] In our present work, however, a question then arises: what serves as the structure-directing agent? As aforementioned, despite the $10^{-6}$ Pa high vacuum condition of TEM chamber, the residual trace amount of oxygen still can induce gradual oxidization of the growing Li nanosheets. To elucidate whether or not oxygen may play the role as surface-directing agent that actuates shape control during anisotropic growth of 2D Li nanostructures, DFT calculations were performed. At first, we calculated the surface energies of Li (111) and (110), which are 34.4 meV/Å$^2$, and 30.9 meV/Å$^2$, respectively. Our calculated results are in very good agreement with the values reported in previous studies.[46] The (111) surface is slightly less energetically favored than the (110) surface, indicating that the basal surface may be easier to be oxidized. In addition, the (111) surface is rougher than the (110) surface at the atomic scale (see Fig. S6), so oxygen molecules may have a larger chance to react with the (111) surface, which could also enhance the oxidation in the basal surface.

Next, we consider oxidized structures of Li slabs. In general, the DFT calculations showed that configurations with undercoordinated oxygen atoms exposed to the surface are less stable than those configurations where the oxygen atom penetrated the surface of the Li slab during relaxation, and was coordinated by 5 or 6 Li atoms. The reaction of oxygen with Li in experiment likely involves oxygen atoms embedding below the top atomic layers of Li, and hence forming a layer of Li oxide at the surface of the particle. To account for the formation of a layer of Li$_2$O, we studied the energetics involved in forming Li$_2$O from Li metal slabs exposing either the (111) or (110) plane. We calculated the oxidation energy $\Delta E$ for the formation of Li$_2$O slabs exposing the low-energy (111) or (110) planes.[28] The oxidation energy per oxygen atom is:

$$\Delta E = [E_{Li2O}^{slab} - N_{Li}E_{Li}^{slab} - N_O E_{O2}/2]/N_O \qquad (2)$$

where $E_{Li2O}^{slab}$ is the energy of the Li$_2$O slab, $N_{Li}$ is the number of Li atoms in the Li$_2$O slab, $N_O$ is the number of oxygen atoms in the Li$_2$O slab, and $E_{O2}$ is the energy of an oxygen molecule, $E_{Li}^{slab}$ is the average energy of a Li atom in the Li (111) or (110) slab; the slab thickness is 12 or 6 atomic layers (~1 nm), respectively. $\Delta E$ is the energy released per oxygen atom in the formation of a Li$_2$O slab from a pristine Li slab reacting with oxygen gas. The results are shown in Figure 3. $\Delta E$ is more negative for Li$_2$O forming from Li slabs exposing the (111) plane, suggesting that the oxidation of the Li (111) surface into a Li$_2$O surface is the energetically favorable process. The Li (111) plane is oxidized to a larger degree than the Li (110) plane.

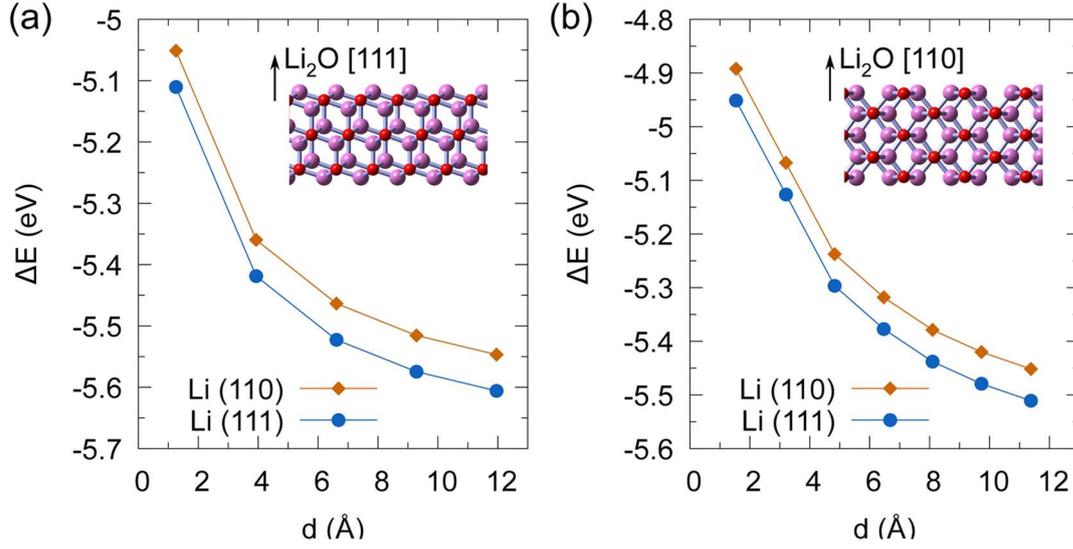

**Figure 3.** The oxidation energy per oxygen atom, $\Delta E$, for the formation of Li$_2$O slabs by oxidation of Li slabs exposing either the (110) or (111) surface, as a function of the thickness $d$ of the resulting Li$_2$O slab. (a) $\Delta E$ for the formation of a Li$_2$O slab exposing the (111) surface. (b) $\Delta E$ for the formation of a Li$_2$O slab exposing the (110) surface.

Unlike the intrinsically layered materials, metal solids with the cubic crystallographic structures are much more difficult to form 2D nanostructures due to the highly symmetric crystal lattice.[29] Therefore, it is only when the cubic symmetry is somehow broken that the 2D anisotropic growth manner can be facilitated. In our present study, such a symmetry-breaking event comes from the preferential oxidization of Li (111) facets due to the residual trace amount (~10$^{-6}$ Pa) of O$_2$. With a combination of in-situ experiments and the DFT calculation results, it is suggested that the formation of Li nanosheets takes place through a mechanistic pathway that is essentially similar to the previously reported cases of CO-assisted synthesis of 2D Pd and Rh nanosheets, where the facet-selective passivation of (111) surfaces by CO molecules was dictating the anisotropic 2D growth process.[5-6] At the early growth stage of nanosheets, isotropic nuclei are usually formed that are defined by lower index surfaces.[17] For the growth of a Li nanosheet, symmetry break induced by preferential oxidization of (111) facets of the nucleus will lead to the formation of a 2D embryo that serve as the real seed for the subsequent anisotropic 2D growth. Specifically, the top and bottom (111) basal planes of 2D Li embryo are confined by the insulating Li$_2$O phase, such that the growth along the [111] direction will be completely suppressed. Under this 2D confined growth model, the newly generated Li atoms will exclusively add to the side surfaces. The continuous lateral extension of the 2D embryos will then give rise to larger nanoplates and further to nanosheets with even larger lateral dimensions. As aforementioned, the lateral extension rate of different Li nanosheets vary from sample to sample. It is only with a slow growth rate that the growth is close to equilibrium and well-faceted nanosheets can be formed, whereas faster growth of the side facets would drive the system out of equilibrium and favor the formation of nanosheets with less-faceted features.[30]

**Plasmonic properties of 2D Li nanosheets**

The availability of free-standing ultrathin Li nanosheets provides us invaluable opportunities to investigate the physical properties of well-defined nanostructures of alkali metals. As free-electron-like metals, alkali metals have low dielectric losses and thereby large plasmon absorption cross-sections, and exhibit surface plasmon resonance (SPR) in the visible wavelength range, like traditional plasmonic metals (Au, Ag and Cu).[31] Due to the extremely high chemical activity of alkali metals, the experimental exploration of the plasmonic properties of alkali metals has been very limited. Early in 1990s, investigations on the plasmon resonances of alkali metal clusters were conducted experimentally.[11, 32] More recently, researchers took advantage of Li plasmonics and realized in operando monitoring of electrochemical evolution of irregular Li metal during battery cycling.[33] The SPR optical properties of plasmonic nanostructures are strongly dependent on their shapes and dimensionalities. In particular, for 2D Au or Ag nanoplates/nanoprisms, they exhibit dimensionality-related plasmonic modes arising from both in-plane and out-of-plane resonances. Here we conducted cathodoluminescence (CL) measurement to probe the plasmonic optical properties of the as-formed Li nanosheets by using an in-situ electrical probing TEM fitted with CL detection systems (see details in Fig. S7).[34] The electron beam was focused on the plasmonic nanostructures to excite plasmon resonances and the radiation of the oscillating plasmon of Li nanosheets contributes to CL which could be collected by the optical fiber near the sample,[35] as schematically illustrated in Fig. 4a. As limited by the electron beam sensitivity of ultrathin Li nanosheets, advanced spatial mapping of CL signal that works well for noble metals, is challenging to achieve on alkali metals, however, we are able to obtain panchromatic CL spectra of Li nanosheets. Figure 4b shows a representative CL spectrum of the in-situ formed Li nanosheet and a strong and broad resonance in the visible wavelength region was observed, which is a result of the overlap of multiple resonant plasmon modes. On the basis of the well-established plasmonic investigations on 2D Au or Ag nanosheets/nanoprisms, the broad plasmon peak of Li nanosheet can be fitted by four Gaussian peaks, and the peaks centred at 713 nm, 614 nm, 545 nm and 446 nm of Li nanosheet in Fig. 4b can be assigned to in-plane dipolar plasmon resonance, in-plane quadrupole resonance, out-of-plane dipole resonance and out-of-plane quadrupole resonance, respectively.[36-38] Although spatial distribution of the different plasmon modes over the whole nanosheet is not achieved, the panchromatic CL spectrum of the ultrathin nanosheet shows clear dimensionality-related plasmonic properties.

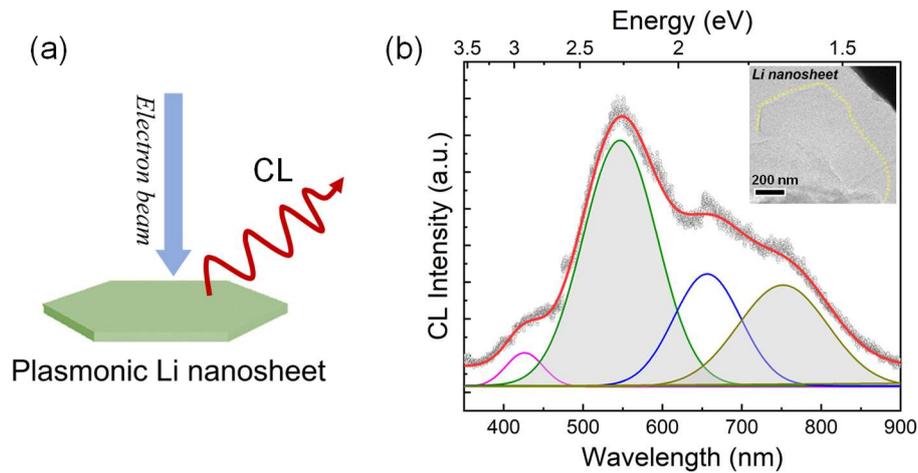

**Figure 4. In-situ CL spectrum measurement from the as-formed Li nanosheet.** (a) Schematic illustration showing CL emission from plasmonic 2D Li nanosheet under the illumination of electron beam. (b) CL spectrum obtained from the as-grown Li nanosheet shown in the inset, which can be well fitted by four Gaussian peaks.

In conclusion, we achieved the in-situ TEM electrochemical growth of ultrathin 2D Li nanosheets by taking advantage of the high vacuum environment inside TEM chamber. Although the presence of the residual trace of $O_2$ (~$10^{-6}$ Pa) within TEM chamber can lead to eventual oxidation of the as-grown Li nanosheets, they, exhibit decent chemical stability enough for real-time monitoring dynamic growth process as well as performing diffraction and spectroscopic analysis of every individual growing nanosheet. The in-situ observation of nanoscale dynamics together with mechanistic insights from DFT calculations provides a coherent picture that the preferential surface passivation of $O_2$ on the (111) surfaces is responsible for suppressing the growth along [111] direction and thus promoting nanosheet formation. The first-ever synthesis of ultrathin Li nanosheets opens up intriguing possibility for exploring fundamental properties of 2D nanostructures of an alkali metal. As a preliminary demonstration, we tested plasmonic optical properties of Li nanosheets via excitation by TEM electrons and observed the dimensionality-related photon emission from the 2D nanosheets. More broadly, our present work also suggests that in-situ TEM electrochemical reactions can provide a unique platform for controlled growth and property studies of nanostructures with well-defined shapes of chemically active metals, such as sodium (Na) and magnesium (Mg). Ongoing work along this line is underway.

**Methods**

**In-situ TEM electrochemical growth.** In-situ characterization was carried out using a PicoFemto in-situ electrical probe holder (provided by ZepTools Technology Company) in a JEOL2010F TEM at 200 kV. The experimental configuration of Li/Li$_2$O/CNT was constructed on the holder in a glovebox in the process described as follows. Metallic Li was scratched by an electrochemically etched tungsten tip in an argon-filled glove box (O$_2$ and H$_2$O content <1 ppm). The gold wire attached with CNT and the tungsten tip with metallic Li were loaded onto the holder inside the glovebox. The holder was then transferred to TEM with the sample exposed to the air for less than 1min, during which the surface of Li metal was oxidized to form a thin layer of Li$_2$O. A piezo-driven nanomanipulator was used to manipulate the tungsten tip to contact a selected CNT inside the TEM column. Electrochemical measurements were conducted using Agilent B2912A under the potentiostatic mode. The design of the experimental setup offers a practical route to achieve solid-state electrochemical reactions and monitor the dynamic process in real-time.[39]

**CL measurement.** CL measurement is based on an electrical-optical TEM holder provided by ZEPTools Technology Company with JEOL 2010F TEM at 200kV in TEM mode. Except for the basic construction of in-situ electrical probe holder to achieve electrochemical growth of ultrathin Li nanosheets, an additional optical fiber of 2 mm in diameter is mounted on the holder near the electrical probe to collect the cathodoluminescence signals from the sample. A grating monochromator iHR 320 from Horiba Inc. and a liquid nitrogen cooled Charge Coupled Device (CCD) were used to record spectra data. When collecting CL spectra, the electron beam was converged to a size slightly smaller than the as-formed Li nanosheet and focused at the center the nanosheet. The collection time was limited within 5 s to avoid detectable radiation damage to Li nanosheet.

## Acknowledgments

We acknowledge the financial support from the Program by Natural Science Foundation (Grant Nos. 21773303, 21872172 and 51421002) of China and Chinese Academy of Sciences (Grant Nos. XDB30000000 and XDB07030100). N.S. acknowledges the Hong Kong Ph.D. Fellowship Scheme. D. P. acknowledges support from the Croucher Foundation through the Croucher Innovation Grant.

## Author contributions

W.W. and X.B conceived the project; M.S. carried out the TEM experiments and analyzed the data under the direction of W.W. and X.B; N.S. and D.P. carried out the DFT calculations; All authors contributed to the discussion of the data and the manuscript.

## Competing interests

The authors declare no competing financial interests.

## References

1. Novoselov, K. S.; Geim, A. K.; Morozov, S. V.; Jiang, D.; Zhang, Y.; Dubonos, S. V.; Grigorieva, I. V.; Firsov, A. A., Electric field effect in atomically thin carbon films. *Science* **2004,** *306* (5696), 666-669.
2. Wang, L.; Zhu, Y.; Wang, J.-Q.; Liu, F.; Huang, J.; Meng, X.; Basset, J.-M.; Han, Y.; Xiao, F.-S., Two-dimensional gold nanostructures with high activity for selective oxidation of carbon-hydrogen bonds. *Nature Communications* **2015,** *6*.
3. Duan, H.; Yan, N.; Yu, R.; Chang, C.-R.; Zhou, G.; Hu, H.-S.; Rong, H.; Niu, Z.; Mao, J.; Asakura, H.; Tanaka, T.; Dyson, P. J.; Li, J.; Li, Y., Ultrathin rhodium nanosheets. *Nature Communications* **2014,** *5*.
4. Chhowalla, M.; Shin, H. S.; Eda, G.; Li, L.-J.; Loh, K. P.; Zhang, H., The chemistry of two-dimensional layered transition metal dichalcogenide nanosheets. *Nature chemistry* **2013,** *5* (4), 263.
5. Huang, X.; Tang, S.; Mu, X.; Dai, Y.; Chen, G.; Zhou, Z.; Ruan, F.; Yang, Z.; Zheng, N., Freestanding palladium nanosheets with plasmonic and catalytic properties. *Nature Nanotechnology* **2011,** *6* (1), 28-32.
6. Zhao, L.; Xu, C.; Su, H.; Liang, J.; Lin, S.; Gu, L.; Wang, X.; Chen, M.; Zheng, N., Single-Crystalline

Rhodium Nanosheets with Atomic Thickness. *Advanced Science* **2015**, *2* (6), 1500100.
7. Yin, A.-X.; Liu, W.-C.; Ke, J.; Zhu, W.; Gu, J.; Zhang, Y.-W.; Yan, C.-H., Ru nanocrystals with shape-dependent surface-enhanced Raman spectra and catalytic properties: controlled synthesis and DFT calculations. *J. Am. Chem. Soc.* **2012,** *134* (50), 20479-20489.
8. Kuang, Y.; Feng, G.; Li, P.; Bi, Y.; Li, Y.; Sun, X., Single-Crystalline Ultrathin Nickel Nanosheets Array from In Situ Topotactic Reduction for Active and Stable Electrocatalysis. *Angewandte Chemie-International Edition* **2016,** *55* (2), 693-697.
9. Amorós, J.; Ravi, S., Correlation among several physicochemical properties of alkali metals in the light of the corresponding states principle. *Phys. Chem. Liq.* **2011,** *49* (1), 9-20.
10. Ellert, C.; Schmidt, M.; Schmitt, C.; Haberland, H.; Guet, C., Reduced oscillator strength in the lithium atom, clusters, and the bulk. *Phys. Rev. B* **1999,** *59* (12), R7841.
11. Bréchignac, C.; Cahuzac, P.; Leygnier, J.; Sarfati, A., Optical response of large lithium clusters: Evolution toward the bulk. *Phys. Rev. Lett.* **1993,** *70* (13), 2036.
12. Huang, J. Y.; Zhong, L.; Wang, C. M.; Sullivan, J. P.; Xu, W.; Zhang, L. Q.; Mao, S. X.; Hudak, N. S.; Liu, X. H.; Subramanian, A.; Fan, H.; Qi, L.; Kushima, A.; Li, J., In situ observation of the electrochemical lithiation of a single SnO(2) nanowire electrode. *Science* **2010,** *330* (6010), 1515.
13. Sun, M.; Qi, K.; Li, X.; Huang, Q.; Wei, J.; Xu, Z.; Wang, W.; Bai, X., Revealing the Electrochemical Lithiation Routes of CuO Nanowires by in Situ TEM. *ChemElectroChem* **2016,** *3* (9), 1296-1300.
14. Ithurria, S.; Bousquet, G.; Dubertret, B., Continuous transition from 3D to 1D confinement observed during the formation of CdSe nanoplatelets. *J. Am. Chem. Soc.* **2011,** *133* (9), 3070-3077.
15. Liu, D. R.; Williams, D. B., The electron-energy-loss spectrum of lithium metal. *Philos. Mag. B-Phys. Condens. Matter Stat. Mech. Electron. Opt. Magn. Prop.* **1986,** *53* (6), L123-L128.
16. Zachman, M. J.; Tu, Z.; Choudhury, S.; Archer, L. A.; Kourkoutis, L. F., Cryo-STEM mapping of solid–liquid interfaces and dendrites in lithium-metal batteries. *Nature* **2018,** *560* (7718), 345.
17. Liao, H.-G.; Zherebetskyy, D.; Xin, H.; Czarnik, C.; Ercius, P.; Elmlund, H.; Pan, M.; Wang, L.-W.; Zheng, H., Facet development during platinum nanocube growth. *Science* **2014,** *345* (6199), 916-919.
18. Xia, Y.; Xia, X.; Peng, H.-C., Shape-controlled synthesis of colloidal metal nanocrystals: thermodynamic versus kinetic products. *J. Am. Chem. Soc.* **2015,** *137* (25), 7947-7966.
19. Xia, X.; Xie, S.; Liu, M.; Peng, H.-C.; Lu, N.; Wang, J.; Kim, M. J.; Xia, Y., On the role of surface diffusion in determining the shape or morphology of noble-metal nanocrystals. *Proceedings of the National Academy of Sciences* **2013,** *110* (17), 6669-6673.
20. Niu, J.; Wang, D.; Qin, H.; Xiong, X.; Tan, P.; Li, Y.; Liu, R.; Lu, X.; Wu, J.; Zhang, T.; Ni, W.; Jin, J., Novel polymer-free iridescent lamellar hydrogel for two-dimensional confined growth of ultrathin gold membranes. *Nature Communications* **2014,** *5*.
21. Huang, X.; Li, S.; Huang, Y.; Wu, S.; Zhou, X.; Li, S.; Gan, C. L.; Boey, F.; Mirkin, C. A.; Zhang, H., Synthesis of hexagonal close-packed gold nanostructures. *Nature Communications* **2011,** *2*.
22. Zhang, H.-R.; Egerton, R. F.; Malac, M., Local thickness measurement through scattering contrast and electron energy-loss spectroscopy. *Micron* **2012,** *43* (1), 8-15.
23. Chee, S. W.; Baraissov, Z.; Loh, N. D.; Matsudaira, P. T.; Mirsaidov, U., Desorption-Mediated Motion of Nanoparticles at the Liquid-Solid Interface. *The Journal of Physical Chemistry C* **2016,** *120* (36), 20462-20470.
24. Wang, J.; Wang, S.-Q., Surface energy and work function of fcc and bcc crystals: Density functional study. *Surf. Sci.* **2014,** *630*, 216-224.
25. Personick, M. L.; Mirkin, C. A., Making sense of the mayhem behind shape control in the synthesis of gold nanoparticles. *J. Am. Chem. Soc.* **2013,** *135* (49), 18238-18247.

26. Lhuillier, E.; Pedetti, S.; Ithurria, S.; Nadal, B.; Heuclin, H.; Dubertret, B., Two-dimensional colloidal metal chalcogenides semiconductors: synthesis, spectroscopy, and applications. *Acc. Chem. Res.* **2015,** *48* (1), 22-30.

27. Luo, L.; Li, Y.; Sun, X.; Li, J.; Hu, E.; Liu, Y.; Tian, Y.; Yang, X.-Q.; Li, Y.; Lin, W.-F., Synthesis and properties of stable sub-2-nm-thick aluminum nanosheets: Oxygen passivation and two-photon luminescence. *Chem* **2020,** *6* (2), 448-459.

28. Radin, M. D.; Rodriguez, J. F.; Tian, F.; Siegel, D. J., Lithium Peroxide Surfaces Are Metallic, While Lithium Oxide Surfaces Are Not. *J. Am. Chem. Soc.* **2012,** *134* (2), 1093-1103.

29. Chen, Y.; Fan, Z.; Zhang, Z.; Niu, W.; Li, C.; Yang, N.; Chen, B.; Zhang, H., Two-dimensional metal nanomaterials: synthesis, properties, and applications. *Chem. Rev.* **2018,** *118* (13), 6409-6455.

30. Ye, X.; Jones, M. R.; Frechette, L. B.; Chen, Q.; Powers, A. S.; Ercius, P.; Dunn, G.; Rotskoff, G. M.; Nguyen, S. C.; Adiga, V. P., Single-particle mapping of nonequilibrium nanocrystal transformations. *Science* **2016,** *354* (6314), 874-877.

31. Blaber, M.; Arnold, M.; Harris, N.; Ford, M.; Cortie, M., Plasmon absorption in nanospheres: A comparison of sodium, potassium, aluminium, silver and gold. *Physica B: Condensed Matter* **2007,** *394* (2), 184-187.

32. Markowicz, P.; Kolwas, K.; Kolwas, M., Experimental determination of free-electron plasma damping rate in large sodium clusters. *Phys. Lett. A* **1997,** *236* (5-6), 543-547.

33. Jin, Y.; Zhou, L.; Yu, J.; Liang, J.; Cai, W.; Zhang, H.; Zhu, S.; Zhu, J., In operando plasmonic monitoring of electrochemical evolution of lithium metal. *Proceedings of the National Academy of Sciences* **2018,** *115* (44), 11168-11173.

34. Yang, S.; Tian, X.; Wang, L.; Wei, J.; Qi, K.; Li, X.; Xu, Z.; Wang, W.; Zhao, J.; Bai, X.; Wang, E., In-situ optical transmission electron microscope study of exciton phonon replicas in ZnO nanowires by cathodoluminescence. *Appl. Phys. Lett.* **2014,** *105* (7).

35. Chaturvedi, P.; Hsu, K. H.; Kumar, A.; Fung, K. H.; Mabon, J. C.; Fang, N. X., Imaging of Plasmonic Modes of Silver Nanoparticles Using High-Resolution Cathodoluminescence Spectroscopy. *Acs Nano* **2009,** *3* (10), 2965-2974.

36. Jin, R. C.; Cao, Y. W.; Mirkin, C. A.; Kelly, K. L.; Schatz, G. C.; Zheng, J. G., Photoinduced conversion of silver nanospheres to nanoprisms. *Science* **2001,** *294* (5548), 1901-1903.

37. Yang, Y.; Matsubara, S.; Xiong, L.; Hayakawa, T.; Nogami, M., Solvothermal synthesis of multiple shapes of silver nanoparticles and their SERS properties. *Journal of Physical Chemistry C* **2007,** *111* (26), 9095-9104.

38. Xiong, Y.; McLellan, J. M.; Chen, J.; Yin, Y.; Li, Z.-Y.; Xia, Y., Kinetically controlled synthesis of triangular and hexagonal nanoplates of palladium and their SPR/SERS properties. *J. Am. Chem. Soc.* **2005,** *127* (48), 17118-17127.

39. Sun, M.; Wei, J.; Xu, Z.; Huang, Q.; Zhao, Y.; Wang, W.; Bai, X., Electrochemical solid-state amorphization in the immiscible Cu-Li system. *Science bulletin* **2018,** *63* (18), 1208-1214.